%% file: main.tex
\documentclass[
]{ceurart}

\usepackage{listings}
\usepackage{booktabs}
\usepackage{subcaption}
\usepackage{float}
\usepackage[]{caption}
\usepackage{adjustbox}
\usepackage{pgfplotstable}
\usepackage{pgfplots}
\usepackage{tikz}
\pgfplotsset{compat=1.18}
\usepackage{numprint}

\newcommand{\Autoref}[1]{%
  \begingroup%
  \def\chapterautorefname{Chapter}%
  \def\sectionautorefname{Section}%
  \def\subsectionautorefname{Section}%
  \def\subsubsectionautorefname{Section}%
  \def\subfigureautorefname{Figure}%
  \autoref{#1}%
  \endgroup%
}

\newcommand{\AM}[1]{\textcolor{black}{#1}}

\begin{document}

\copyrightyear{2024}
\copyrightclause{Copyright for this paper by its authors.
  Use permitted under Creative Commons License Attribution 4.0
  International (CC BY 4.0).}

\conference{LIRAI’24: 2nd Legal Information Retrieval meets Artificial Intelligence Workshop co-located with the 35th ACM Hypertext Conference, September 10, 2024, Poznan, Poland}

\title{Finding Needles in Emb(a)dding Haystacks: Legal Document Retrieval via Bagging and SVR Ensembles}

\author[1]{Kevin B{\"o}nisch}[%
orcid=0009-0005-6962-2646,
email=k.boenisch@outlook.com,
url=https://www.texttechnologylab.org/team/kevin-boenisch/,
]
\cormark[1]

\author[1]{Alexander Mehler}[%
orcid=0000-0003-2567-7539,
email=mehler@em.uni-frankfurt.de,
url=https://www.texttechnologylab.org/team/alexander-mehler/,
]

\address[1]{Text Technology Lab, Goethe University Frankfurt, Germany }

\cortext[1]{Corresponding author.}

\begin{abstract}
    We introduce 
    a 
    retrieval approach leveraging \textit{Support Vector Regression} (SVR) ensembles, bootstrap aggregation (bagging), and embedding spaces on the \textit{German Dataset for Legal Information Retrieval} (GerDaLIR).
    By conceptualizing the retrieval task in terms of multiple binary needle-in-a-haystack subtasks, we show improved recall over the baselines (0.849 $>$ 0.803 | 0.829) using our voting ensemble,
    suggesting promising initial results, without training or fine-tuning any deep learning models. 
    %
    Our approach holds potential for further enhancement, particularly through refining the encoding models and optimizing hyperparameters.
\end{abstract}

\begin{keywords}
  legal information retrieval \sep
  support vector regression \sep
  word embeddings \sep
  bagging ensemble
\end{keywords}

\maketitle

\section{Introduction}

Information Retrieval (IR) fundamentally involves the process of identifying and extracting 
information units from an extensive collection, often in response to a specific query \cite{Salton:McGill:1983}.
Legal Information Retrieval (LIR) represents the application of this concept within the legal domain, wherein the document collection consists of relevant judicial passages \cite{Sansone:Sperli:2022}. 
Among these documents, some are particularly relevant to a given legal case (query), necessitating the task of identifying and retrieving these specific documents.
Consequently, a variety of Natural Language Processing (NLP) methodologies, such as \textit{term frequency-inverse document frequency} (TF-IDF) \citep{Salton:1989} (cf.\ \cite{wehnert:2021:legalnorm} for applying this in the legal domain) and \textit{BM25} ranking \citep{Manning_Raghavan_Schütze_2008}, are utilized to rank and compare collections of documents in order to retrieve the most pertinent ones.
With the advent of the transformer architecture \citep{vaswani2023attentionneed}, language modeling has become a crucial component of LIR as well \citep{nogueira2020passagererankingbert}.
In this context, pre-trained bidirectional transformer models such as \textit{BERT} \citep{devlin2019bertpretrainingdeepbidirectional}, \textit{RoBERTa} \citep{liu2019robertarobustlyoptimizedbert} and \textit{DeBERTa} \citep{he2021debertadecodingenhancedbertdisentangled} are especially 
valuable, as they contextually capture 
semantic features of potentially relevant documents through high-dimensional embedding vectors, making them distinguishable and comparable on a computational level.
Through fine-tuning, these models can be further adapted to more domain-specific downstream tasks such as LIR, as exemplified by \textit{LEGAL-BERT} \citep{chalkidis2020legalbertmuppetsstraightlaw} and \textit{DISC-LawLLM} \citep{yue2023disclawllmfinetuninglargelanguage}.
Despite the significant potential these models offer for many NLP tasks \citep{brown2020languagemodelsfewshotlearners}, their lack of explainability remains a widespread criticism \citep{luo2024understandingutilizationsurveyexplainability, zhao2023explainabilitylargelanguagemodels}, which imparts them with a ''black box`` character that is critically scrutinized, particularly in the legal domain.
\newline
In order to effectively train and apply these models, essential datasets tailored for (LIR) include those from the \textit{Legal Information Extraction and Entailment} (COLIE) competition \citep{colie}, \textit{SigmaLaw} \citep{sugathadasa2018legaldocumentretrievalusing}, \textit{FIRE 2017 IRLED} \citep{Mandal2017OverviewOT}, and \textit{GerDaLIR} \citep{wrzalik-krechel-2021-gerdalir}.
To our knowledge, \textit{GerDaLIR} stands as the sole German dataset for LIR, highlighting the predominance of English datasets and the absence of non-English retrieval models.
\newline
To address this, we present 
results of our retrieval technique on the German \textit{GerDaLIR} dataset, which surpass all baseline models. Our approach ensembles weak \textit{Support Vector Regressors} (SVR) \citep{svr}, embeddings from a variety of pre-trained bidirectional transformers and employs \textit{Bagging} \citep{Breiman:1996zz} techniques parallel to those used in \textit{Random Forests} \citep{Breiman2001}.
In doing so, we frame the LIR problem akin to a needle-in-a-haystack task, as detailed in \Autoref{sec:haystack}.
Our results demonstrate an increase in recall (\textbf{0.849} $>$ 0.803 | 0.829) over the best baseline models, achieved without the need for fine-tuning any large deep learning models.
Finally, we publish our soure code on GitHub\footnote{\url{https://github.com/TheItCrOw/lirai24}}.

\section{Related Work}

For Information Retrieval (IR) in general, TF-IDF stands as one of the primary methodologies \citep{Salton:Buckley:1988,Beel2016}, owing to its simplicity and adaptability to various corpora and information sources.
Consequently, within the aforementioned COLIEE competition, TF-IDF was widely used to directly retrieve or pre-process and pre-rank a given collection of documents \citep{nguyen2020jnlpteamdeeplearning, coliee2}.
Following this, various BERT variants are employed for re-ranking \citep{nogueira2020passagererankingbert} or downstream task fine-tuning \citep{chalkidis2020legalbertmuppetsstraightlaw}, often in combination with methodologies such as TF-IDF. 
Additionally, contextualized embeddings from bidirectional pre-trained transformers like BERT \citep{kamalloo2023evaluatingembeddingapisinformation, galkeembeddings2017} are used for re-ranking or leveraging closeness metrics, such as cosine similarity, to fetch relevant documents.
Finally, to enhance these models, feature engineering is applied to introduce potentially unique features, such as adding metadata to the documents, incorporating external data, or using model ensembles \citep{wehnert:2021:legalnorm}.
In the following sections, we will outline in detail the relevant technologies used for our retrieval model.

\subsection{Embeddings}

Transforming words and documents into a non-textual representation is a crucial task for nearly any NLP application. 
Among various methods, the 
vector space model introduced by Salton \citep{1975vectorspace,Salton:1989} is arguably the most relevant and influential approach for achieving this.
As a result, much effort has been invested in projecting textual data into high-dimensional spaces to capture semantic and contextual features. 
This trend was exemplified in 2014 with \textit{Word2Vec} \citep{mikolov2013efficientestimationwordrepresentations} and has since evolved with more recent models such as BERT and DeBERTa.
These models generate high-dimensional embedding spaces for linguistic units, enabling spatial Euclidean measurements to capture semantic similarities and facilitate contextual retrieval or model training.

\subsection{Supported Vector Machine Regression}

Support Vector Machines (SVM) are a popular machine learning tool for classification and regression (SVR), introduced by \citep{svr-origin} in 1995. However, to our knowledge, they have not been widely used for LIR and IR in general. 
Akin to linear regression, Support Vector Machines (SVMs) aim to find the optimal hyperplane that effectively separates the two classes in the data. This hyperplane maximizes the margin, which is the distance between the hyperplane and the closest data points from each class, known as \textit{support vectors}.
SVR (Support Vector Regression) utilizes a similar concept where the margin is defined as an error tolerance of the model, known as the $\epsilon$-insensitive tube \citep{svr-origin}. 
Given training data $(\mathbf{x}_i, y_i)$ for $i = 1, \ldots, n$, SVR seeks to find a function $f(\mathbf{x}) = \mathbf{w} \cdot \mathbf{x} + b$ that approximates the true values $y_i$ with minimal error, subject to the following constraints outlined in \citep[p.~153 onward]{svr-origin} and \citep[p.~157]{drucker:svr:1996}:
\newline
\[
y_i - (\mathbf{w} \cdot \mathbf{x}_i + b) \leq \epsilon + \xi_i
\]
\[
(\mathbf{w} \cdot \mathbf{x}_i + b) - y_i \leq \epsilon + \xi_i
\]
\newline
where $\xi_i \geq 0$ are slack variables representing the deviation from the margin, and $\epsilon$ controls the width of the tube. The objective is to minimize the following regularized error function:
\newline
\[
\frac{1}{2} \| \mathbf{w} \|^2 + C \sum_{i=1}^{n} (\xi_i + \xi_i^*)
\]
\newline
subject to the constraints above, where $C$ is a regularization parameter that balances the trade-off between the margin and the training error. The dual problem formulation involves computing Lagrange multipliers $\alpha_i, \alpha_i^*$ for each constraint, leading to the dual problem:
\newline
\[
\min_{\alpha, \alpha^*} \frac{1}{2} \sum_{i,j} (\alpha_i - \alpha_i^*)(\alpha_j - \alpha_j^*) \mathbf{x}_i \cdot \mathbf{x}_j + \epsilon \sum_{i} (\alpha_i + \alpha_i^*) - \sum_{i} (y_i (\alpha_i - \alpha_i^*))
\]
\newline
with constraints $\sum_{i} (\alpha_i - \alpha_i^*) = 0$ and $0 \leq \alpha_i, \alpha_i^* \leq C$ for $i = 1, \ldots, n$. The solution $\mathbf{w}$ to the primal problem can be expressed in terms of the support vectors $\mathbf{x}_i$ and their corresponding Lagrange multipliers $\alpha_i - \alpha_i^*$.
\newline
In practice, SVR effectively handles nonlinear relationships through kernel functions, mapping the input space into a higher-dimensional feature space where a linear model is constructed.

\subsection{Bagging}

Bootstrap Aggregating (Bagging), introduced by \citep{Breiman:1996zz} in 1996, is an ensemble learning technique designed to enhance model accuracy by combining multiple base learners which also provides the capability to partition the problem into smaller sub-problems.
The key aspects of bagging involve:

\begin{itemize}
    \item Creating multiple subsets of the original dataset, thereby forming smaller sub-problems for base learners.
    \item Training each base learner independently on these subsets.
    \item Combining the predictions of the individual learners in the ensemble to make a final prediction.
\end{itemize}

The strength of bagging lies in its ability to utilize an ensemble of smaller models \citep{SHARAFATI20213521}, each trained on a subset of the dataset, particularly effective in scenarios with large feature sets and class imbalance. In such cases, overfitting can be a potential concern, and bagging can mitigate this risk effectively and hence improve generalization \citep{qian2024baggingimprovesgeneralizationexponentially}.

\subsection{GerDaLIR}

GerDaLIR \citep{wrzalik-krechel-2021-gerdalir} is a \textbf{G}erman \textbf{D}ataset for \textbf{L}egal \textbf{I}nformation \textbf{R}etrieval based on the Open Legal Data platform.
The dataset consists of 123,000 queries and 131,000 case documents, which are segmented into over 3 million passages. Each query is labeled with at least one document and hence multiple passages.
The provided task is a precedent retrieval task based on case documents from the \textit{Open Legal Data}\footnote{\url{https://openlegaldata.io/}} platform.
The authors provide several baseline models for LIR, including TF-IDF, BM25, and deep learning approaches such as BERT re-ranking \citep{nogueira2020passagererankingbert}.

\section{Methodology}

In this section, we outline our approach to LIR on the GerDaLIR dataset, following the structure of a data science project report from start to finish, to demonstrate our findings and the chain of thought that led to our conclusions and final models.

\input{tex/eda-fig}

\subsection{Explorative Data Analysis}\label{sec:eda}

We begin by experimenting with various encoder transformer models. 
The intuition is to determine whether a given query is projected next to or at least very close to relevant documents in the collection embedding space, allowing for the possibility of retrieving documents based on their location in this space.
Figure 1a displays excerpts of such a space 
as t-SNE \citep{Maaten:Hinton:2008} plots. 
We tested a total of eight different models and placed exemplary query embeddings into the collection space to observe their relation to 
relevant passages.
It can be seen that while none of the queries inserted into the collection space are positioned directly adjacent to their relevant document passages, the relevant passages are often in close proximity to the queries -- this is consistent with the approach of Salton's vector space model \cite{Salton:1989}.
This is particularly evident when examining the \textit{longformer} model \citep{beltagy2020longformerlongdocumenttransformer}, which appears to generate the most effective clustering for this task. 
The \textit{longformer} was specifically trained to handle longer text sequences beyond lexical tokens 
or (short) sentences, which benefits the handling of queries and passages that typically range from two to six sentences in length.
\newline
Another observation 
is that the embedding space is 
heavily influenced by the length of the passage or query. 
Texts of the same length, regardless of their semantic similarity, tended to be placed close together, while contextually more relevant texts were often placed further apart.
We therefore chunked the passages into texts of equal length, which resulted in a better placement of relevant passages in relation to each other, as shown in Figure 1b.
With regard to the LIR technique presented in this paper, we have not yet utilised this knowledge as our initial focus is on establishing a baseline. However, we intend to incorporate this approach in future iterations.

\begin{figure}
    \centering
    \includegraphics[width=\textwidth]{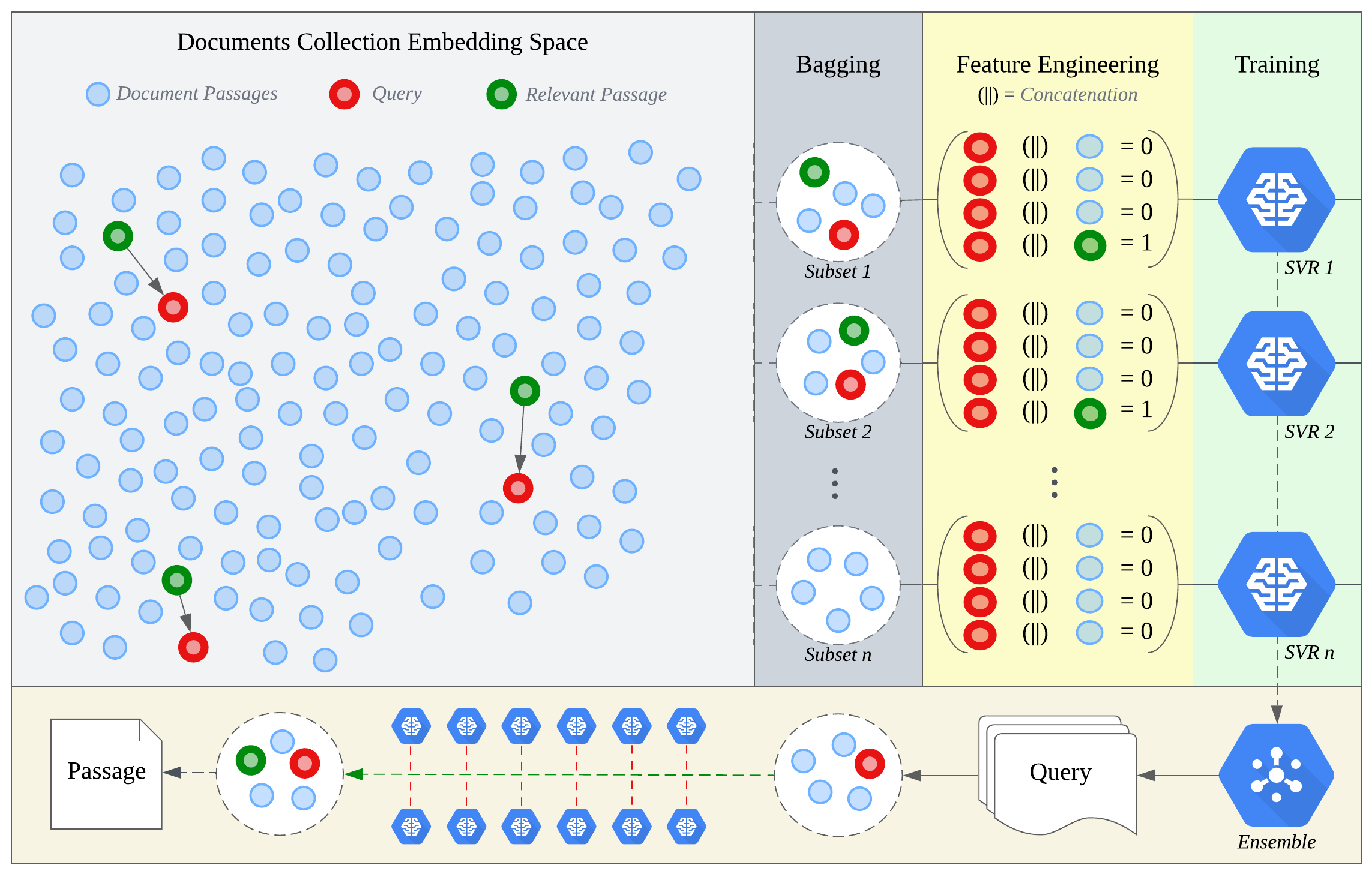}
    \caption{Modeling the 
    flow of the needle-in-a-haystack training, we begin by partitioning the 
    document space into several subsets (bagging), with each subset being assigned a separate SVR model for training.
     For each query in a subset, we identify the top $k$ nearest passages through their vector spaces and concatenate their embeddings into a single feature embedding. Consequently, each query is associated with $k - 1$ negative labels and one positive label. 
     The SVR model is trained to find this single positive label within the haystack. This process is repeated for each subset and query. 
     During prediction, each model in the group predicts a match for its respective subset. If only one model recognizes a positive match, the corresponding section is marked as relevant and output.}
    \label{fig:architecture}
\end{figure}

\subsection{Finding the Needle in a Haystack}\label{sec:haystack}

Similar to other re-ranking techniques \citep{nogueira2020passagererankingbert}, we use the embeddings generated by the longformer model, as it seemed to work best for our task, to formulate the LIR procedure as outlined in \Autoref{fig:architecture}.\footnote{\AM{For the use of embedding methods in retrieval see also \cite{Karpukhin:et:al:2020}. See \cite{Zhu:Yuan:et:al:2024} for a recent survey of embedding methods, especially with respect to large language models in IR.}}
\newline
First, we generate an embedding for each passage, resulting in a $3.095.383 \times 768$ embedding matrix. This forms our collection embedding space.
Next, we employ bagging and partition the embedding space into $s$ subsets with some overlapping, where each subset is used to train its own SVR model, utilizing the Radial Basis Function (RBF) kernel. For this baseline, we partitioned it into $35$ subsets with $60\%$ overlapping, resulting in a final model ensemble of 35 SVRs.
For each query, we generate a $1 \times 768$ embedding and place it in the embedding space. Then, we collect the $k$ nearest neighbors, which for this baseline 
is $50$. Our training assumes that the relevant passage to be retrieved is somewhere among these $k$ passages.
\newline
We then solve a binary regression training task for all of these $k$ passages, each with $k-1$ negative labels and one positive label (the relevant passage among the $k$). 
As features for the SVR model, we concatenate the query's embedding vector with each passage, resulting in a $k \times (2 \times 768)$ feature matrix for the query. This process is repeated for each query in the subset and for each subset in the collection.
This results in a $\numprint{5034360} \times \numprint{1536}$ feature matrix for training. 
We then normalize the features with \textit{scikit}'s \textsc{StandardScaler}\footnote{See \url{https://scikit-learn.org/stable/modules/generated/sklearn.preprocessing.StandardScaler.html}; this is just a $z$-score transformation.}, 
before splitting it 
according to a $0.9/0.1$ train-test split. The GerDaLIR dataset includes a separate test split on which our baseline is evaluated, allowing for a very large training split. 
\newline
For training, we utilize the \textsc{RAPIDS cuML}\footnote{\url{https://github.com/rapidsai/cuml}} implementation of SVR to leverage the GPU resources, even for classical machine learning such as SVMs. Training was conducted on two NVIDIA Quadro RTX GPUs, each with 48GB of RAM. The entire process of calculating embeddings, performing bagging, creating the feature space, normalizing, and training took a total of two days, requiring 200GB of RAM swap. To enhance efficiency, we downscaled the precision of the matrices to 32-bit floats.

\subsection{Results}
\definecolor{lightblue}{rgb}{0.8,0.85,1}
\begin{table}[h]
    \centering
    \begin{tabular}{|l|c|c|c|c|c|}
        \hline
        \rowcolor{lightgray}
        \textbf{Method} & \textbf{Mode} & \textbf{MRR@10} & \textbf{nDCG@20} & \textbf{Recall@100} & \textbf{Recall@1000} \\
        \hline
        TF-IDF & P & 0.333 & 0.375 & 0.651 & 0.768 \\
               & D & 0.336 & 0.386 & 0.701 & 0.809 \\
        BM25 (k1=1.20, b=0.75) & P & 0.365 & 0.409 & 0.693 & 0.800 \\
                               & D & 0.386 & 0.434 & 0.734 & 0.827 \\
        \rowcolor{lightblue}
        BM25 tuned (k1=0.51, b=0.72) & P & 0.372 & 0.417 & 0.703 & \textbf{0.803} \\
        BM25 tuned (k1=0.90, b=0.98) & D & 0.391 & 0.439 & 0.737 & \textbf{0.829} \\
        WCS - GloVe & P & 0.242 & 0.278 & 0.539 & 0.695 \\
                    & D & 0.134 & 0.166 & 0.420 & 0.625 \\
        WCS - fastText & P & 0.257 & 0.295 & 0.582 & 0.726 \\
                       & D & 0.153 & 0.188 & 0.468 & 0.668 \\
        Neural Re-ranking - BERT & P & 0.416 & 0.465 & \textbf{0.745} & 0.789 \\
        Neural Re-ranking - ELECTRA & P & \textbf{0.436} & \textbf{0.481} & \textbf{0.743} & 0.789 \\
        \hline
        \rowcolor{lightgray}
        \textbf{Method} & \textbf{Mode} & \multicolumn{4}{c|}{\textbf{Recall}} \\
        \rowcolor{yellow}
        \textbf{SVR Ensemble} & \textbf{P} & \multicolumn{4}{c|}{\textbf{0.849}} \\
        \hline
    \end{tabular}
    \caption{The baseline measures of the GerDaLIR dataset are compared to our preliminary LIR technique. In this context, Mode P and D refer to passage-wise and document-wise retrieval, respectively. Metrics such as MRR and nDCG have not yet been evaluated on our ensemble. The original authors utilized Recall@100 and Recall@1000 for re-ranking, a restriction that we do not impose. The results indicate that our preliminary ensemble surpasses all baseline measures.}
    \label{tab:results}
\end{table}

\input{tex/results}

\Autoref{tab:results} presents the results of the GerDaLIR dataset baselines alongside our ensemble. In addition to recall metrics, further details of our training results are outlined in \Autoref{tab:results-2} and \ref{tab:results-3}.
The seemingly perfect scores for class $0$ are a result of the nature of the needle-in-a-haystack task and should not be misinterpreted as indicating perfect model accuracy. Given that we sample, on average, $k-1$ negative labels for each positive label, the precision and recall for this highly imbalanced class are distorted.
Therefore, it is necessary to focus primarily on the results for class $1$, where it is evident that even with a relatively small $k$ of just $50$, our ensemble consistently identifies relevant passages with high recall, precision, accuracy, and F1 score.
From this point of view, we now outline our decision to select the SVR of all ML models as our ensemble models:

\begin{itemize}
    \item The ever-present danger with high-dimensional data is overfitting \citep{tsuda2023benignoverfittingnonsparsehighdimensional}. 
    This risk is exacerbated by the fact that for each query we sample, on average, $k-1$ negative samples with only one positive sample. 
    One way to mitigate such overfitting is to utilize L1 or L2 regularization \citep{l1l2}, which SVR implements by default.
    \item The SVR approach involves mapping data into a higher-dimensional space than the original data set, in order to achieve better separability. From the beginning, we found that concatenating embedding vectors as features is favorable, making SVR a very suitable model for our purposes.
    \item Because of kernel functions such as RBF, SVRs are able to capture nonlinear relationships \citep{svrsurvey}. This capability is crucial for the needle-in-a-haystack task at hand, as a linear function would not be sufficient to separate the two classes.
    But this is also a reason why overfitting is made easier. A variety of nonlinear functions should therefore be experimented with here.
    \item The RAPIDS implementation gave us access to traditional stochastic machine learning models that run on GPUs, which greatly accelerated the training process. Given the large feature space, this step was critical.
\end{itemize}

Finally, as previously mentioned, SVRs are significantly more interpretable than deep learning models, providing greater transparency.

\section{Conclusion and Future Work}

We showcased a novel approach for LIR, combining several machine learning methods such as SVR, bagging, and embedding spaces.
While these initial results are promising, we need to work on the following areas:
\begin{enumerate}
    \item As demonstrated by our exploratory data analysis in \Autoref{sec:eda}, the embedding spaces heavily depend on the length of the texts. 
    If we were to further partition the passages and queries into texts of equal length, the overall ensemble and retrieval process could benefit from it. Further investigation into this approach is necessary.
    Additionally, our approach is based on the observation that relevant passages are generally proximate to the queries within the embedding space. Nevertheless, as illustrated in Figure 1a, there are instances where outlier passages are positioned far from their corresponding queries, making it exceedingly difficult for our ensemble model to identify them. 
    This is one of the key constraints that we need to address, particularly through the investigation of the impact of increasing the value of $k$.
    \item Due to time and hardware constraints, we had to opt for a relatively small $k=50$ for first level retrieval.
    Increasing the radius of the initial retrieval layer, within which the models search for relevant information, could significantly enhance recall.
    \item The initial idea was to utilize multiple embedding spaces generated by various encoder models and concatenate them, thereby further increasing the feature space by higher dimensions. This approach has proven effective in several high-profile NLP competitions\footnote{\url{https://www.kaggle.com/competitions/feedback-prize-english-language-learning/discussion/369457}}. For these preliminary results, we opted for a single embedding model, but we aim to expand this approach.
\end{enumerate}
Finally, we recognize the need for more German-based models and approaches, particularly within the field of law. The encoding models currently employed primarily process English text and rarely specialize in legal domains. While fine-tuned versions like LEGAL-BERT exist, they predominantly cater to non-German texts.
The development of a German pre-trained encoder model, akin to LEGAL-BERT, could help bridge this gap and presents a promising avenue for future research.
\newline
Our present approach combines SVMs with transformer-based embedding models to develop an appropriate feature space for the classification task at hand. 
The very simple method evolves as a process in which better and better embedding models can improve the second main part of our method, just as more efficient classification methods can improve the classical ML part that makes up the first main part of our approach.
The obvious question as to why we do not rely directly on fine-tuning a pretrained embedding model lies in the transparency that the SVM-based approach offers us as a classification tool. 
This transparency is necessary because we divide the original document space into overlapping subspaces that compete, so to speak, for the processing of a query by the classifiers assigned to them.
Thus, we divide the overall task into a series of subtasks, each of which can be transparently processed by classifiers, where the partitioning of the overall document space can reflect the specific topological conditions of the underlying embedding space, providing additional transparency. 
However, by using more and more information about the topology of the document space, the classifiers can adapt more and more efficiently to its structure, eventually becoming less and less complex (below the level of RBFs used so far).
In this way, we envision an approach that takes advantage of both the ongoing process of inventing ever more expressive embedding spaces and models for topology preserving segmentation of such spaces as a prerequisite for training ever more efficient classifiers. 
This may eventually contribute to a kind of information retrieval that combines transparency with expressiveness in terms of numerical representation of documents -- very much in the spirit of Salton's classical vector space model.

\bibliography{citations}

\appendix

\end{document}

%% file: tex/eda-fig.tex
\begin{figure}[htp]
    \centering
    \resizebox{\textwidth}{!}{
    \begin{tabular}{@{}c@{\hspace{0.1cm}}c@{\hspace{0.1cm}}c@{}}
        \begin{subfigure}[b]{0.3\textwidth}
            \centering
            \includegraphics[width=\textwidth, height=4cm]{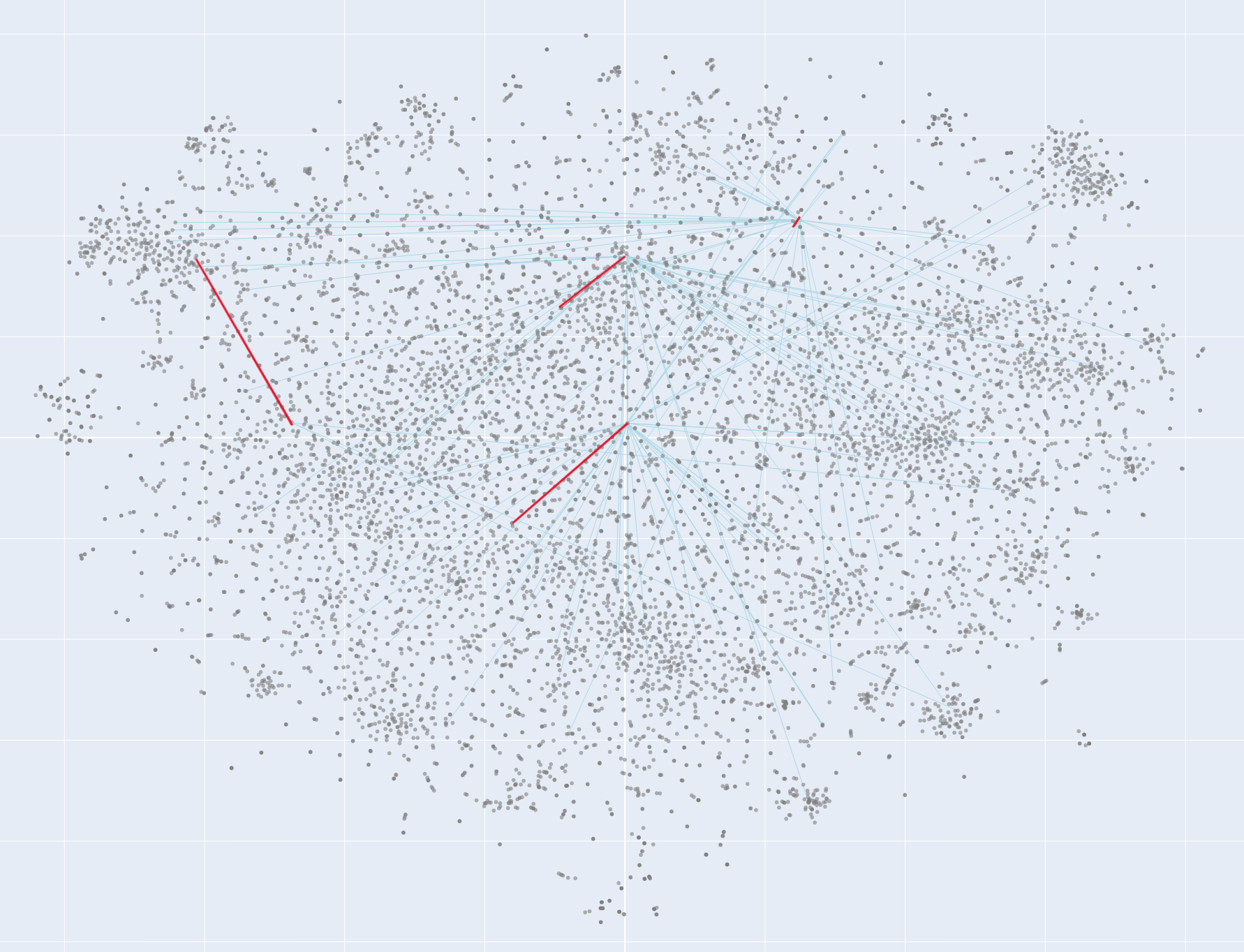}
            \caption*{\scriptsize DistilRoBERTa}
        \end{subfigure} &
        \begin{subfigure}[b]{0.3\textwidth}
            \centering
            \includegraphics[width=\textwidth, height=4cm]{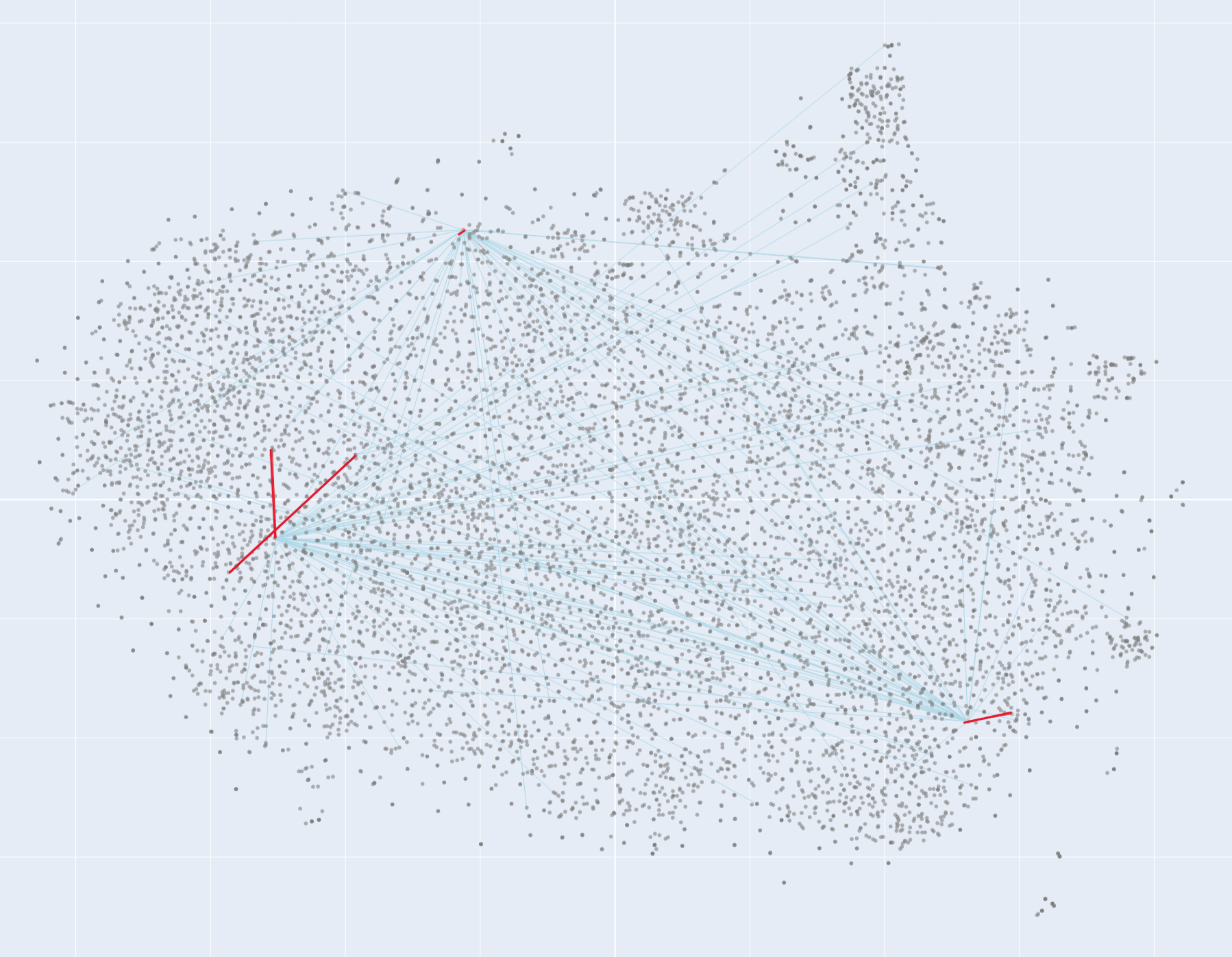}
            \caption*{\scriptsize RoBERTa\_base}
        \end{subfigure} &
        \begin{subfigure}[b]{0.3\textwidth}
            \centering
            \includegraphics[width=\textwidth, height=4cm]{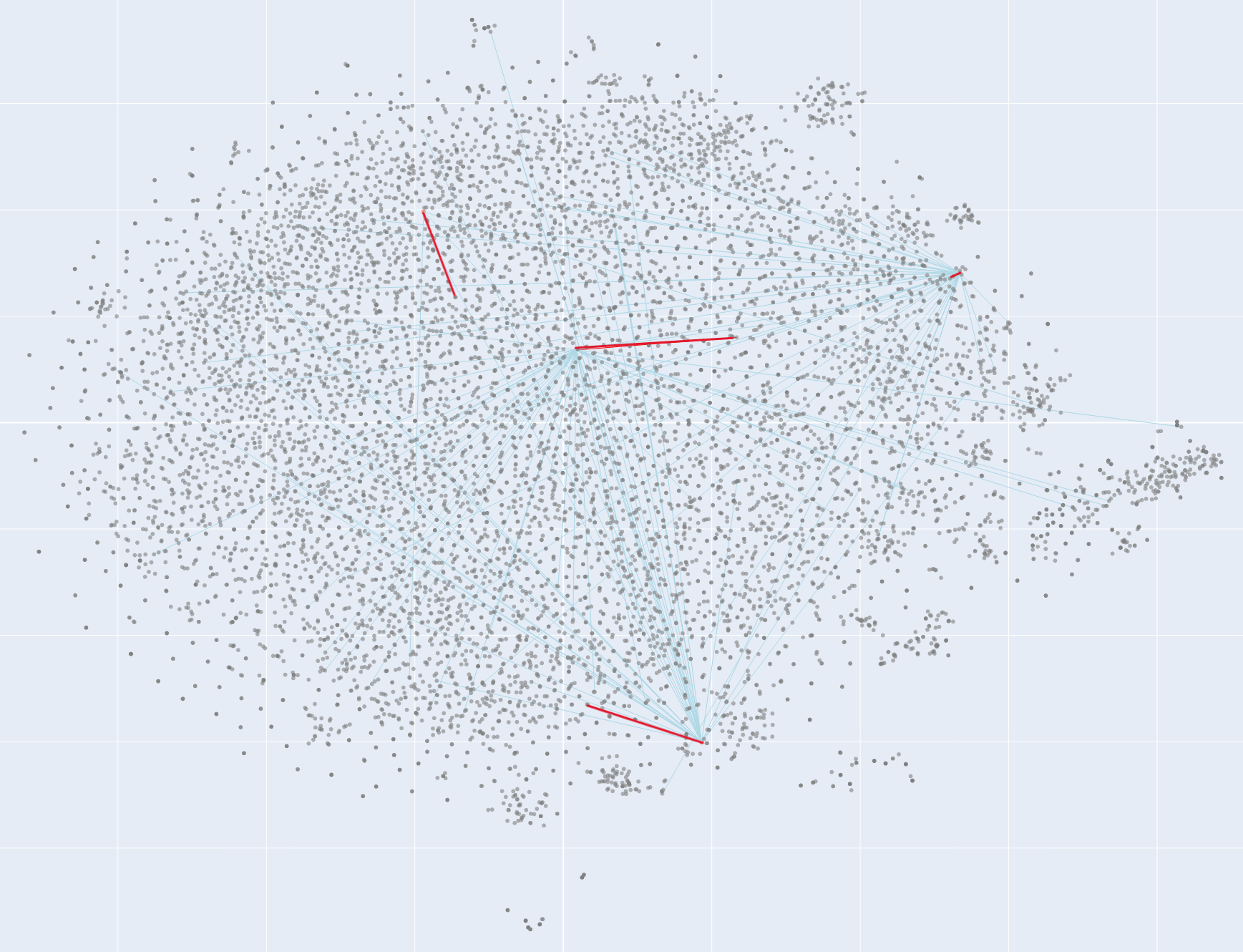}
            \caption*{\scriptsize RoBERTa\_large}
        \end{subfigure} \\
        \begin{subfigure}[b]{0.3\textwidth}
            \centering
            \includegraphics[width=\textwidth, height=4cm]{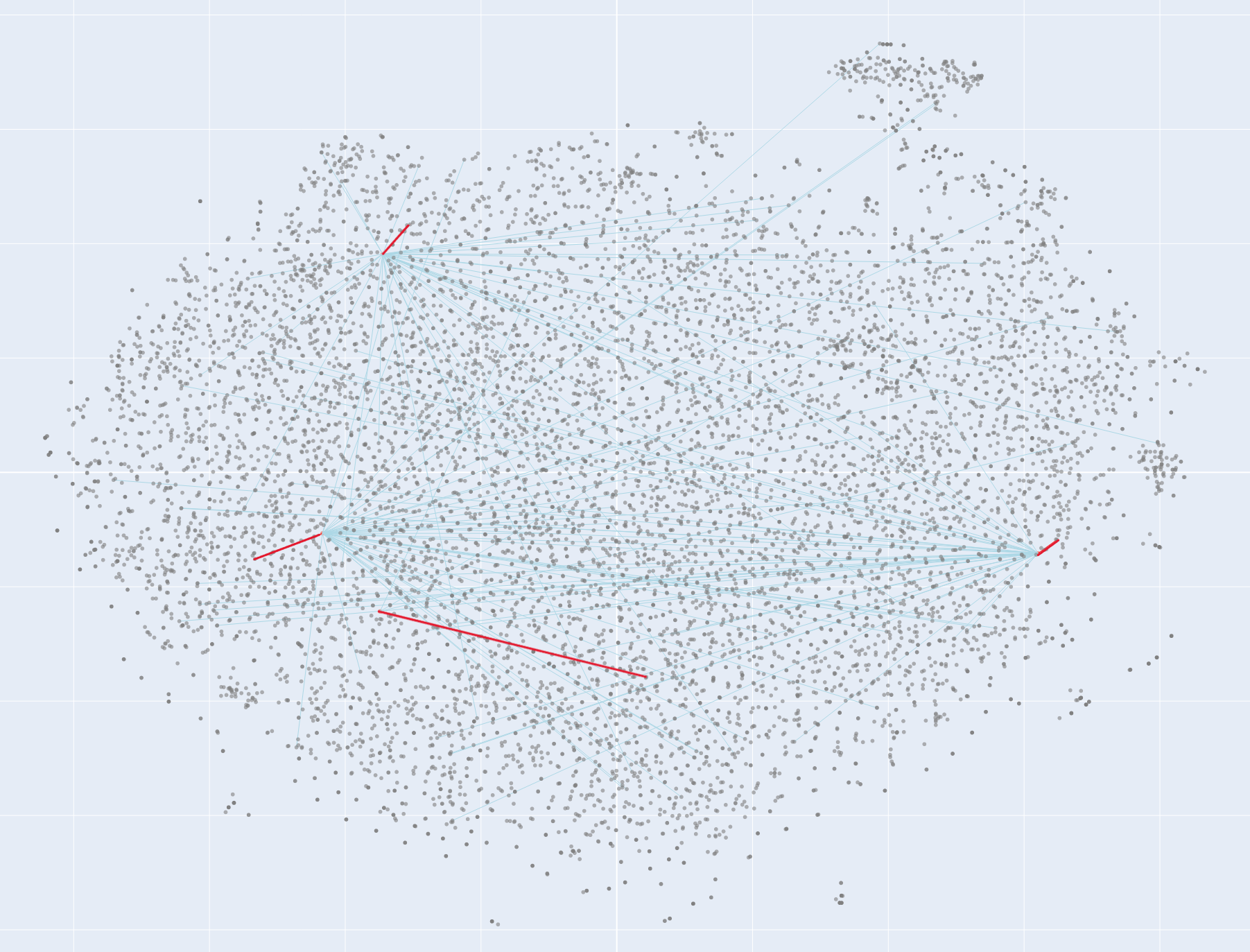}
            \caption*{\scriptsize DeBERTa\_v3\_large}
        \end{subfigure} &
        \begin{subfigure}[b]{0.3\textwidth}
            \centering
            \includegraphics[width=\textwidth, height=4cm]{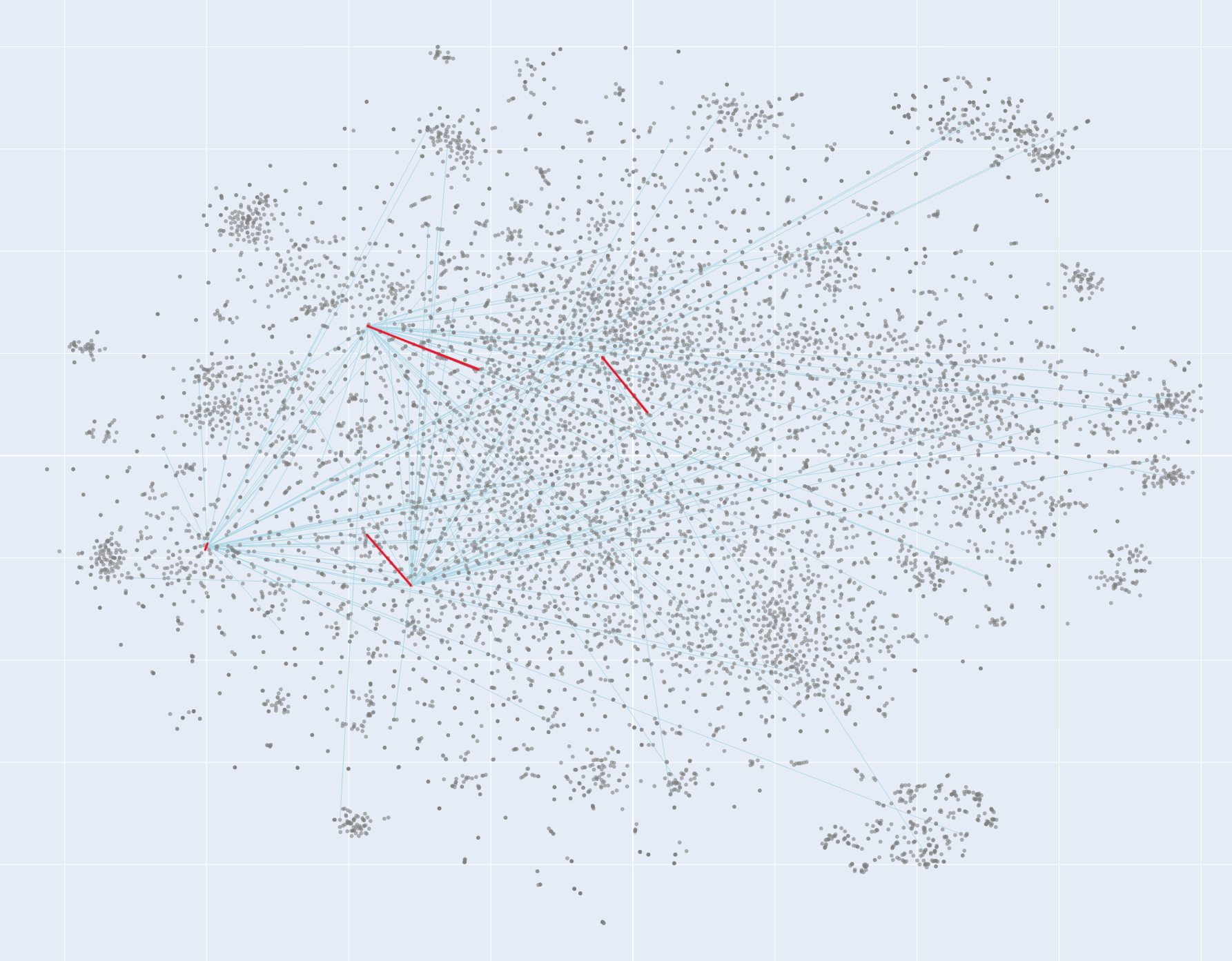}
            \caption*{\scriptsize DistilBERT}
        \end{subfigure} &
        \begin{subfigure}[b]{0.3\textwidth}
            \centering
            \includegraphics[width=\textwidth, height=4cm]{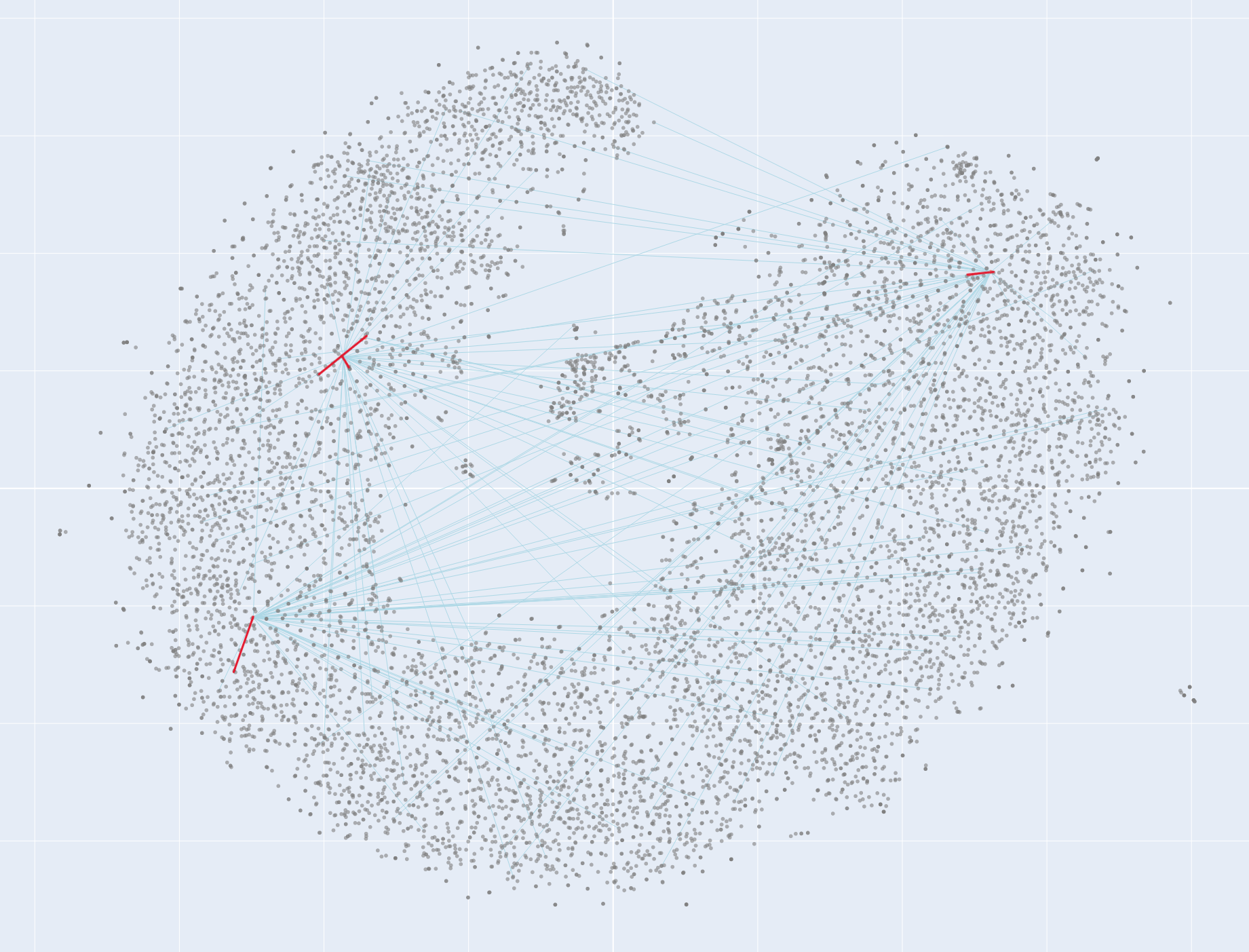}
            \caption*{\scriptsize  Longformer\_base}
        \end{subfigure}
    \end{tabular}}
    \subcaption*{\footnotesize \textbf{Figure 1a:} 
    t-SNE plots of embedding spaces from different encoder models, generated for an exemplary sample of 10,000 collection passages. Each grey point represents a document passage in the collection. The red lines indicate the shortest distance from placed queries to their labelled relevant passages. Four queries were placed in each collection space, showing different distances to their relevant passages. While none of the relevant passages were the closest to their respective queries, it can be observed that in many cases the relevant passage is in close proximity. In particular, the \textit{longformer\_base} embedding space seems to capture the context best, as the entire collection forms a U-shaped cluster, and the distances from query to passage remain consistently the smallest, observable by the very short and barely visible red lines.}
    \label{fig:eda-1}
    
    \vspace{0.3cm}
    
    \begin{subfigure}[b]{0.496\linewidth}
        \centering
        \includegraphics[width=\linewidth]{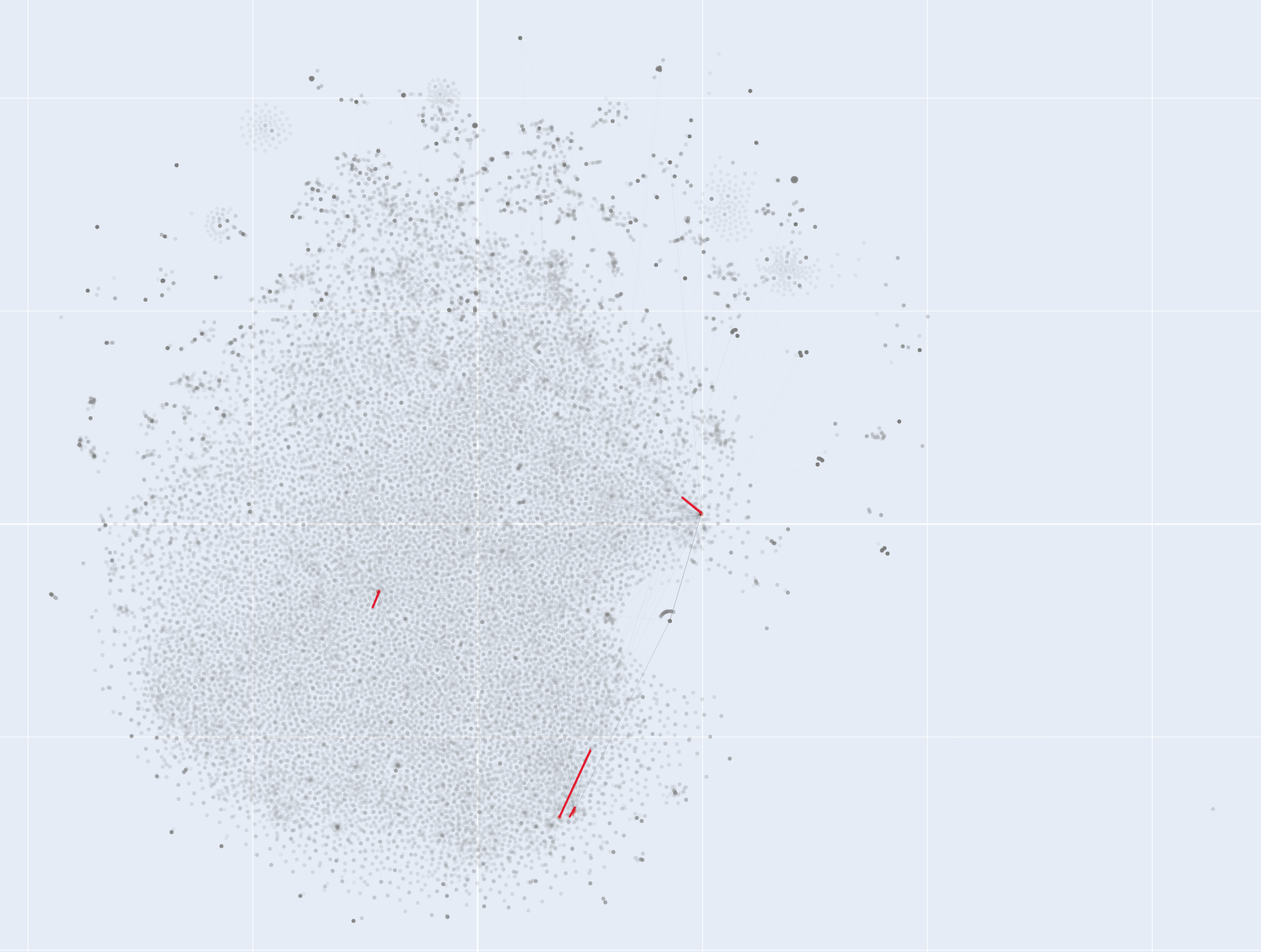}
        \caption*{\footnotesize deBERTa\_base chunked}
    \end{subfigure}
    \hfill
    \begin{subfigure}[b]{0.496\linewidth}
        \centering
        \includegraphics[width=\linewidth]{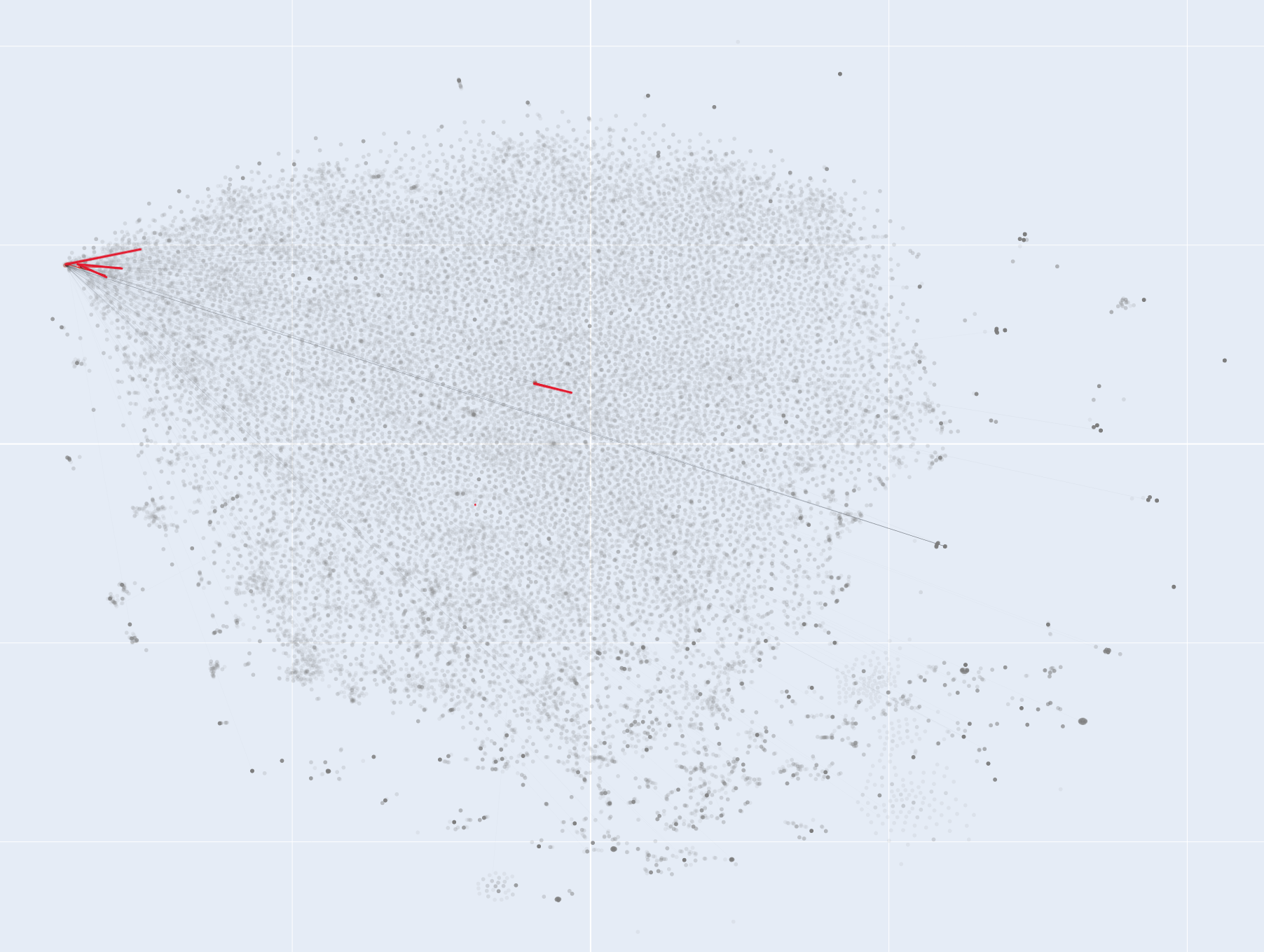}
        \caption*{\footnotesize Longformer\_base chunked}
    \end{subfigure}
    \subcaption*{\footnotesize \textbf{Figure 1b:} Exemplary t-SNE plots of the same document passages shown in Figure 1a, with the addition of chunking the passages into segments of similar lengths matching their overall average length, 
    show that the average distance to the most relevant passage of each query is smaller compared to the spaces without chunking.}
\end{figure}

%% file: tex/results.tex
\begin{figure}[h!]
  \centering
  \begin{minipage}[t]{0.34\textwidth}
    \caption{Model Accuracy}
    \begin{tabular}{|l|c|}
    \hline
    \textbf{Metric} & \textbf{Value} \\
    \hline
    Test set accuracy & 0.9966 \\
    \hline
    Precision & 0.9987 \\
    \rowcolor{lightblue}
    \hline
    Recall & 0.8527 \\
    \rowcolor{lightblue}
    \hline
    F1-score & 0.9199 \\
    \hline
    \end{tabular}
    \label{tab:results-2}
  \end{minipage}
  \hfill
  \begin{minipage}[t]{0.64\textwidth}
    \centering
    \caption{Classification Report}
    \begin{tabular}{|l|c|c|c|c|}
    \hline
    \textbf{Class} & \textbf{Precision} & \textbf{Recall} & \textbf{F1-score} & \textbf{Support} \\
    \hline
    0 & 1.00 & 1.00 & 1.00 & 491915 \\
    \rowcolor{lightblue}
    \hline
    1 & 1.00 & 0.85 & 0.92 & 11521 \\
    \hline
    \textbf{Accuracy} & \multicolumn{4}{c|}{1.00 (503436 samples)} \\
    \hline
    \textbf{Macro avg} & 1.00 & 0.93 & 0.96 & 503436 \\
    \hline
    \textbf{Weighted avg} & 1.00 & 1.00 & 1.00 & 503436 \\
    \hline
    \end{tabular}
    \label{tab:results-3}
  \end{minipage}
\end{figure}

%% file: main.bbl
\begin{thebibliography}{41}
\expandafter\ifx\csname natexlab\endcsname\relax\def\natexlab#1{#1}\fi
\providecommand{\url}[1]{\texttt{#1}}
\providecommand{\href}[2]{#2}
\providecommand{\path}[1]{#1}
\providecommand{\DOIprefix}{doi:}
\providecommand{\ArXivprefix}{arXiv:}
\providecommand{\URLprefix}{URL: }
\providecommand{\Pubmedprefix}{pmid:}
\providecommand{\doi}[1]{\href{http://dx.doi.org/#1}{\path{#1}}}
\providecommand{\Pubmed}[1]{\href{pmid:#1}{\path{#1}}}
\providecommand{\bibinfo}[2]{#2}
\ifx\xfnm\relax \def\xfnm[#1]{\unskip,\space#1}\fi
\bibitem[{Salton and McGill(1983)}]{Salton:McGill:1983}
\bibinfo{author}{G.~Salton}, \bibinfo{author}{M.~McGill},
\newblock \bibinfo{title}{Introduction to modern information retrieval},
\newblock \bibinfo{year}{1983}. \URLprefix \url{https://api.semanticscholar.org/CorpusID:43685115}.
\bibitem[{Sansone and Sperl{\'\i}(2022)}]{Sansone:Sperli:2022}
\bibinfo{author}{C.~Sansone}, \bibinfo{author}{G.~Sperl{\'\i}},
\newblock \bibinfo{title}{Legal information retrieval systems: State-of-the-art and open issues},
\newblock \bibinfo{journal}{Information Systems} \bibinfo{volume}{106} (\bibinfo{year}{2022}) \bibinfo{pages}{101967}.
\bibitem[{Salton(1989)}]{Salton:1989}
\bibinfo{author}{G.~Salton}, \bibinfo{title}{Automatic Text Processing: The Transformation, Analysis, and Retrieval of Information by Computer}, \bibinfo{publisher}{Addison Wesley}, \bibinfo{address}{Reading, Massachusetts}, \bibinfo{year}{1989}.
\bibitem[{Wehnert et~al.(2021)Wehnert, Sudhi, Dureja, Kutty, Shahania, and De~Luca}]{wehnert:2021:legalnorm}
\bibinfo{author}{S.~Wehnert}, \bibinfo{author}{V.~Sudhi}, \bibinfo{author}{S.~Dureja}, \bibinfo{author}{L.~Kutty}, \bibinfo{author}{S.~Shahania}, \bibinfo{author}{E.~W. De~Luca},
\newblock \bibinfo{title}{Legal norm retrieval with variations of the bert model combined with tf-idf vectorization},
\newblock in: \bibinfo{booktitle}{Proceedings of the Eighteenth International Conference on Artificial Intelligence and Law}, ICAIL '21, \bibinfo{publisher}{Association for Computing Machinery}, \bibinfo{address}{New York, NY, USA}, \bibinfo{year}{2021}, p. \bibinfo{pages}{285–294}. \URLprefix \url{https://doi.org/10.1145/3462757.3466104}. \DOIprefix\doi{10.1145/3462757.3466104}.
\bibitem[{Manning et~al.(2008)Manning, Raghavan, and Schütze}]{Manning_Raghavan_Schütze_2008}
\bibinfo{author}{C.~D. Manning}, \bibinfo{author}{P.~Raghavan}, \bibinfo{author}{H.~Schütze}, \bibinfo{title}{Introduction to Information Retrieval}, \bibinfo{publisher}{Cambridge University Press}, \bibinfo{year}{2008}.
\bibitem[{Vaswani et~al.(2023)Vaswani, Shazeer, Parmar, Uszkoreit, Jones, Gomez, Kaiser, and Polosukhin}]{vaswani2023attentionneed}
\bibinfo{author}{A.~Vaswani}, \bibinfo{author}{N.~Shazeer}, \bibinfo{author}{N.~Parmar}, \bibinfo{author}{J.~Uszkoreit}, \bibinfo{author}{L.~Jones}, \bibinfo{author}{A.~N. Gomez}, \bibinfo{author}{L.~Kaiser}, \bibinfo{author}{I.~Polosukhin}, \bibinfo{title}{Attention is all you need}, \bibinfo{year}{2023}. \URLprefix \url{https://arxiv.org/abs/1706.03762}. \href{http://arxiv.org/abs/1706.03762}{{\tt arXiv:1706.03762}}.
\bibitem[{Nogueira and Cho(2020)}]{nogueira2020passagererankingbert}
\bibinfo{author}{R.~Nogueira}, \bibinfo{author}{K.~Cho}, \bibinfo{title}{Passage re-ranking with bert}, \bibinfo{year}{2020}. \URLprefix \url{https://arxiv.org/abs/1901.04085}. \href{http://arxiv.org/abs/1901.04085}{{\tt arXiv:1901.04085}}.
\bibitem[{Devlin et~al.(2019)Devlin, Chang, Lee, and Toutanova}]{devlin2019bertpretrainingdeepbidirectional}
\bibinfo{author}{J.~Devlin}, \bibinfo{author}{M.-W. Chang}, \bibinfo{author}{K.~Lee}, \bibinfo{author}{K.~Toutanova}, \bibinfo{title}{Bert: Pre-training of deep bidirectional transformers for language understanding}, \bibinfo{year}{2019}. \URLprefix \url{https://arxiv.org/abs/1810.04805}. \href{http://arxiv.org/abs/1810.04805}{{\tt arXiv:1810.04805}}.
\bibitem[{Liu et~al.(2019)Liu, Ott, Goyal, Du, Joshi, Chen, Levy, Lewis, Zettlemoyer, and Stoyanov}]{liu2019robertarobustlyoptimizedbert}
\bibinfo{author}{Y.~Liu}, \bibinfo{author}{M.~Ott}, \bibinfo{author}{N.~Goyal}, \bibinfo{author}{J.~Du}, \bibinfo{author}{M.~Joshi}, \bibinfo{author}{D.~Chen}, \bibinfo{author}{O.~Levy}, \bibinfo{author}{M.~Lewis}, \bibinfo{author}{L.~Zettlemoyer}, \bibinfo{author}{V.~Stoyanov}, \bibinfo{title}{Roberta: A robustly optimized bert pretraining approach}, \bibinfo{year}{2019}. \URLprefix \url{https://arxiv.org/abs/1907.11692}. \href{http://arxiv.org/abs/1907.11692}{{\tt arXiv:1907.11692}}.
\bibitem[{He et~al.(2021)He, Liu, Gao, and Chen}]{he2021debertadecodingenhancedbertdisentangled}
\bibinfo{author}{P.~He}, \bibinfo{author}{X.~Liu}, \bibinfo{author}{J.~Gao}, \bibinfo{author}{W.~Chen}, \bibinfo{title}{Deberta: Decoding-enhanced bert with disentangled attention}, \bibinfo{year}{2021}. \URLprefix \url{https://arxiv.org/abs/2006.03654}. \href{http://arxiv.org/abs/2006.03654}{{\tt arXiv:2006.03654}}.
\bibitem[{Chalkidis et~al.(2020)Chalkidis, Fergadiotis, Malakasiotis, Aletras, and Androutsopoulos}]{chalkidis2020legalbertmuppetsstraightlaw}
\bibinfo{author}{I.~Chalkidis}, \bibinfo{author}{M.~Fergadiotis}, \bibinfo{author}{P.~Malakasiotis}, \bibinfo{author}{N.~Aletras}, \bibinfo{author}{I.~Androutsopoulos}, \bibinfo{title}{Legal-bert: The muppets straight out of law school}, \bibinfo{year}{2020}. \URLprefix \url{https://arxiv.org/abs/2010.02559}. \href{http://arxiv.org/abs/2010.02559}{{\tt arXiv:2010.02559}}.
\bibitem[{Yue et~al.(2023)Yue, Chen, Wang, Li, Shen, Liu, Zhou, Xiao, Yun, Huang, and Wei}]{yue2023disclawllmfinetuninglargelanguage}
\bibinfo{author}{S.~Yue}, \bibinfo{author}{W.~Chen}, \bibinfo{author}{S.~Wang}, \bibinfo{author}{B.~Li}, \bibinfo{author}{C.~Shen}, \bibinfo{author}{S.~Liu}, \bibinfo{author}{Y.~Zhou}, \bibinfo{author}{Y.~Xiao}, \bibinfo{author}{S.~Yun}, \bibinfo{author}{X.~Huang}, \bibinfo{author}{Z.~Wei}, \bibinfo{title}{Disc-lawllm: Fine-tuning large language models for intelligent legal services}, \bibinfo{year}{2023}. \URLprefix \url{https://arxiv.org/abs/2309.11325}. \href{http://arxiv.org/abs/2309.11325}{{\tt arXiv:2309.11325}}.
\bibitem[{Brown et~al.(2020)Brown, Mann, Ryder, Subbiah, Kaplan, Dhariwal, Neelakantan, Shyam, Sastry, Askell, Agarwal, Herbert-Voss, Krueger, Henighan, Child, Ramesh, Ziegler, Wu, Winter, Hesse, Chen, Sigler, Litwin, Gray, Chess, Clark, Berner, McCandlish, Radford, Sutskever, and Amodei}]{brown2020languagemodelsfewshotlearners}
\bibinfo{author}{T.~B. Brown}, \bibinfo{author}{B.~Mann}, \bibinfo{author}{N.~Ryder}, \bibinfo{author}{M.~Subbiah}, \bibinfo{author}{J.~Kaplan}, \bibinfo{author}{P.~Dhariwal}, \bibinfo{author}{A.~Neelakantan}, \bibinfo{author}{P.~Shyam}, \bibinfo{author}{G.~Sastry}, \bibinfo{author}{A.~Askell}, \bibinfo{author}{S.~Agarwal}, \bibinfo{author}{A.~Herbert-Voss}, \bibinfo{author}{G.~Krueger}, \bibinfo{author}{T.~Henighan}, \bibinfo{author}{R.~Child}, \bibinfo{author}{A.~Ramesh}, \bibinfo{author}{D.~M. Ziegler}, \bibinfo{author}{J.~Wu}, \bibinfo{author}{C.~Winter}, \bibinfo{author}{C.~Hesse}, \bibinfo{author}{M.~Chen}, \bibinfo{author}{E.~Sigler}, \bibinfo{author}{M.~Litwin}, \bibinfo{author}{S.~Gray}, \bibinfo{author}{B.~Chess}, \bibinfo{author}{J.~Clark}, \bibinfo{author}{C.~Berner}, \bibinfo{author}{S.~McCandlish}, \bibinfo{author}{A.~Radford}, \bibinfo{author}{I.~Sutskever}, \bibinfo{author}{D.~Amodei}, \bibinfo{title}{Language models are few-shot learners}, \bibinfo{year}{2020}. \URLprefix
  \url{https://arxiv.org/abs/2005.14165}. \href{http://arxiv.org/abs/2005.14165}{{\tt arXiv:2005.14165}}.
\bibitem[{Luo and Specia(2024)}]{luo2024understandingutilizationsurveyexplainability}
\bibinfo{author}{H.~Luo}, \bibinfo{author}{L.~Specia}, \bibinfo{title}{From understanding to utilization: A survey on explainability for large language models}, \bibinfo{year}{2024}. \URLprefix \url{https://arxiv.org/abs/2401.12874}. \href{http://arxiv.org/abs/2401.12874}{{\tt arXiv:2401.12874}}.
\bibitem[{Zhao et~al.(2023)Zhao, Chen, Yang, Liu, Deng, Cai, Wang, Yin, and Du}]{zhao2023explainabilitylargelanguagemodels}
\bibinfo{author}{H.~Zhao}, \bibinfo{author}{H.~Chen}, \bibinfo{author}{F.~Yang}, \bibinfo{author}{N.~Liu}, \bibinfo{author}{H.~Deng}, \bibinfo{author}{H.~Cai}, \bibinfo{author}{S.~Wang}, \bibinfo{author}{D.~Yin}, \bibinfo{author}{M.~Du}, \bibinfo{title}{Explainability for large language models: A survey}, \bibinfo{year}{2023}. \URLprefix \url{https://arxiv.org/abs/2309.01029}. \href{http://arxiv.org/abs/2309.01029}{{\tt arXiv:2309.01029}}.
\bibitem[{Rabelo et~al.(2021)Rabelo, Kim, Goebel, Yoshioka, Kano, and Satoh}]{colie}
\bibinfo{author}{J.~Rabelo}, \bibinfo{author}{M.-Y. Kim}, \bibinfo{author}{R.~Goebel}, \bibinfo{author}{M.~Yoshioka}, \bibinfo{author}{Y.~Kano}, \bibinfo{author}{K.~Satoh},
\newblock \bibinfo{title}{Coliee 2020: Methods for legal document retrieval and entailment},
\newblock in: \bibinfo{editor}{N.~Okazaki}, \bibinfo{editor}{K.~Yada}, \bibinfo{editor}{K.~Satoh}, \bibinfo{editor}{K.~Mineshima} (Eds.), \bibinfo{booktitle}{New Frontiers in Artificial Intelligence}, \bibinfo{publisher}{Springer International Publishing}, \bibinfo{address}{Cham}, \bibinfo{year}{2021}, pp. \bibinfo{pages}{196--210}.
\bibitem[{Sugathadasa et~al.(2018)Sugathadasa, Ayesha, de~Silva, Perera, Jayawardana, Lakmal, and Perera}]{sugathadasa2018legaldocumentretrievalusing}
\bibinfo{author}{K.~Sugathadasa}, \bibinfo{author}{B.~Ayesha}, \bibinfo{author}{N.~de~Silva}, \bibinfo{author}{A.~S. Perera}, \bibinfo{author}{V.~Jayawardana}, \bibinfo{author}{D.~Lakmal}, \bibinfo{author}{M.~Perera}, \bibinfo{title}{Legal document retrieval using document vector embeddings and deep learning}, \bibinfo{year}{2018}. \URLprefix \url{https://arxiv.org/abs/1805.10685}. \href{http://arxiv.org/abs/1805.10685}{{\tt arXiv:1805.10685}}.
\bibitem[{Mandal et~al.(2017)Mandal, Ghosh, Bhattacharya, Pal, and Ghosh}]{Mandal2017OverviewOT}
\bibinfo{author}{A.~Mandal}, \bibinfo{author}{K.~Ghosh}, \bibinfo{author}{A.~Bhattacharya}, \bibinfo{author}{A.~Pal}, \bibinfo{author}{S.~Ghosh},
\newblock \bibinfo{title}{Overview of the fire 2017 irled track: Information retrieval from legal documents},
\newblock in: \bibinfo{booktitle}{Fire}, \bibinfo{year}{2017}. \URLprefix \url{https://api.semanticscholar.org/CorpusID:39265594}.
\bibitem[{Wrzalik and Krechel(2021)}]{wrzalik-krechel-2021-gerdalir}
\bibinfo{author}{M.~Wrzalik}, \bibinfo{author}{D.~Krechel},
\newblock \bibinfo{title}{{G}er{D}a{LIR}: A {G}erman dataset for legal information retrieval},
\newblock in: \bibinfo{editor}{N.~Aletras}, \bibinfo{editor}{I.~Androutsopoulos}, \bibinfo{editor}{L.~Barrett}, \bibinfo{editor}{C.~Goanta}, \bibinfo{editor}{D.~Preotiuc-Pietro} (Eds.), \bibinfo{booktitle}{Proceedings of the Natural Legal Language Processing Workshop 2021}, \bibinfo{publisher}{Association for Computational Linguistics}, \bibinfo{address}{Punta Cana, Dominican Republic}, \bibinfo{year}{2021}, pp. \bibinfo{pages}{123--128}. \URLprefix \url{https://aclanthology.org/2021.nllp-1.13}. \DOIprefix\doi{10.18653/v1/2021.nllp-1.13}.
\bibitem[{Hearst et~al.(1998)Hearst, Dumais, Osuna, Platt, and Scholkopf}]{svr}
\bibinfo{author}{M.~Hearst}, \bibinfo{author}{S.~Dumais}, \bibinfo{author}{E.~Osuna}, \bibinfo{author}{J.~Platt}, \bibinfo{author}{B.~Scholkopf},
\newblock \bibinfo{title}{Support vector machines},
\newblock \bibinfo{journal}{IEEE Intelligent Systems and their Applications} \bibinfo{volume}{13} (\bibinfo{year}{1998}) \bibinfo{pages}{18--28}. \DOIprefix\doi{10.1109/5254.708428}.
\bibitem[{Breiman(1996)}]{Breiman:1996zz}
\bibinfo{author}{L.~Breiman},
\newblock \bibinfo{title}{{Bagging Predictors}},
\newblock \bibinfo{journal}{Machine Learning} \bibinfo{volume}{24} (\bibinfo{year}{1996}) \bibinfo{pages}{123--140}. \DOIprefix\doi{10.1007/BF00058655}.
\bibitem[{Breiman(2001)}]{Breiman2001}
\bibinfo{author}{L.~Breiman},
\newblock \bibinfo{title}{Random forests},
\newblock \bibinfo{journal}{Machine Learning} \bibinfo{volume}{45} (\bibinfo{year}{2001}) \bibinfo{pages}{5--32}. \URLprefix \url{https://doi.org/10.1023/A:1010933404324}. \DOIprefix\doi{10.1023/A:1010933404324}.
\bibitem[{Salton and Buckley(1988)}]{Salton:Buckley:1988}
\bibinfo{author}{G.~Salton}, \bibinfo{author}{C.~Buckley},
\newblock \bibinfo{title}{Term weighting approaches in automatic text retrieval},
\newblock \bibinfo{journal}{Information Processing Management} \bibinfo{volume}{24} (\bibinfo{year}{1988}) \bibinfo{pages}{513--523}.
\bibitem[{Beel et~al.(2016)Beel, Gipp, Langer, and Breitinger}]{Beel2016}
\bibinfo{author}{J.~Beel}, \bibinfo{author}{B.~Gipp}, \bibinfo{author}{S.~Langer}, \bibinfo{author}{C.~Breitinger},
\newblock \bibinfo{title}{Research-paper recommender systems: a literature survey},
\newblock \bibinfo{journal}{International Journal on Digital Libraries} \bibinfo{volume}{17} (\bibinfo{year}{2016}) \bibinfo{pages}{305--338}. \URLprefix \url{https://doi.org/10.1007/s00799-015-0156-0}. \DOIprefix\doi{10.1007/s00799-015-0156-0}.
\bibitem[{Nguyen et~al.(2020)Nguyen, Vuong, Nguyen, Dang, Bui, Vu, Nguyen, Tran, Satoh, and Nguyen}]{nguyen2020jnlpteamdeeplearning}
\bibinfo{author}{H.-T. Nguyen}, \bibinfo{author}{H.-Y.~T. Vuong}, \bibinfo{author}{P.~M. Nguyen}, \bibinfo{author}{B.~T. Dang}, \bibinfo{author}{Q.~M. Bui}, \bibinfo{author}{S.~T. Vu}, \bibinfo{author}{C.~M. Nguyen}, \bibinfo{author}{V.~Tran}, \bibinfo{author}{K.~Satoh}, \bibinfo{author}{M.~L. Nguyen}, \bibinfo{title}{Jnlp team: Deep learning for legal processing in coliee 2020}, \bibinfo{year}{2020}. \URLprefix \url{https://arxiv.org/abs/2011.08071}. \href{http://arxiv.org/abs/2011.08071}{{\tt arXiv:2011.08071}}.
\bibitem[{Kim et~al.(2023)Kim, Rabelo, Goebel, Yoshioka, Kano, and Satoh}]{coliee2}
\bibinfo{author}{M.-Y. Kim}, \bibinfo{author}{J.~Rabelo}, \bibinfo{author}{R.~Goebel}, \bibinfo{author}{M.~Yoshioka}, \bibinfo{author}{Y.~Kano}, \bibinfo{author}{K.~Satoh},
\newblock \bibinfo{title}{Coliee 2022 summary: Methods for legal document retrieval and entailment},
\newblock in: \bibinfo{editor}{Y.~Takama}, \bibinfo{editor}{K.~Yada}, \bibinfo{editor}{K.~Satoh}, \bibinfo{editor}{S.~Arai} (Eds.), \bibinfo{booktitle}{New Frontiers in Artificial Intelligence}, \bibinfo{publisher}{Springer Nature Switzerland}, \bibinfo{address}{Cham}, \bibinfo{year}{2023}, pp. \bibinfo{pages}{51--67}.
\bibitem[{Kamalloo et~al.(2023)Kamalloo, Zhang, Ogundepo, Thakur, Alfonso-Hermelo, Rezagholizadeh, and Lin}]{kamalloo2023evaluatingembeddingapisinformation}
\bibinfo{author}{E.~Kamalloo}, \bibinfo{author}{X.~Zhang}, \bibinfo{author}{O.~Ogundepo}, \bibinfo{author}{N.~Thakur}, \bibinfo{author}{D.~Alfonso-Hermelo}, \bibinfo{author}{M.~Rezagholizadeh}, \bibinfo{author}{J.~Lin}, \bibinfo{title}{Evaluating embedding apis for information retrieval}, \bibinfo{year}{2023}. \URLprefix \url{https://arxiv.org/abs/2305.06300}. \href{http://arxiv.org/abs/2305.06300}{{\tt arXiv:2305.06300}}.
\bibitem[{Galke et~al.(2017)Galke, Saleh, and Scherp}]{galkeembeddings2017}
\bibinfo{author}{L.~Galke}, \bibinfo{author}{A.~Saleh}, \bibinfo{author}{A.~Scherp}, \bibinfo{title}{Word embeddings for practical information retrieval}, \bibinfo{howpublished}{INFORMATIK 2017}, \bibinfo{year}{2017}. \DOIprefix\doi{10.18420/in2017_215}.
\bibitem[{Salton et~al.(1975)Salton, Wong, and Yang}]{1975vectorspace}
\bibinfo{author}{G.~Salton}, \bibinfo{author}{A.~Wong}, \bibinfo{author}{C.~S. Yang},
\newblock \bibinfo{title}{A vector space model for automatic indexing},
\newblock \bibinfo{journal}{Commun. ACM} \bibinfo{volume}{18} (\bibinfo{year}{1975}) \bibinfo{pages}{613–620}. \URLprefix \url{https://doi.org/10.1145/361219.361220}. \DOIprefix\doi{10.1145/361219.361220}.
\bibitem[{Mikolov et~al.(2013)Mikolov, Chen, Corrado, and Dean}]{mikolov2013efficientestimationwordrepresentations}
\bibinfo{author}{T.~Mikolov}, \bibinfo{author}{K.~Chen}, \bibinfo{author}{G.~Corrado}, \bibinfo{author}{J.~Dean}, \bibinfo{title}{Efficient estimation of word representations in vector space}, \bibinfo{year}{2013}. \URLprefix \url{https://arxiv.org/abs/1301.3781}. \href{http://arxiv.org/abs/1301.3781}{{\tt arXiv:1301.3781}}.
\bibitem[{Vapnik(1995)}]{svr-origin}
\bibinfo{author}{V.~N. Vapnik}, \bibinfo{title}{The Nature of Statistical Learning Theory}, \bibinfo{publisher}{Springer, New York}, \bibinfo{year}{1995}.
\bibitem[{Drucker et~al.(1996)Drucker, Burges, Kaufman, Smola, and Vapnik}]{drucker:svr:1996}
\bibinfo{author}{H.~Drucker}, \bibinfo{author}{C.~J.~C. Burges}, \bibinfo{author}{L.~Kaufman}, \bibinfo{author}{A.~Smola}, \bibinfo{author}{V.~Vapnik},
\newblock \bibinfo{title}{Support vector regression machines},
\newblock in: \bibinfo{editor}{M.~Mozer}, \bibinfo{editor}{M.~Jordan}, \bibinfo{editor}{T.~Petsche} (Eds.), \bibinfo{booktitle}{Advances in Neural Information Processing Systems}, volume~\bibinfo{volume}{9}, \bibinfo{publisher}{MIT Press}, \bibinfo{year}{1996}. \URLprefix \url{https://proceedings.neurips.cc/paper_files/paper/1996/file/d38901788c533e8286cb6400b40b386d-Paper.pdf}.
\bibitem[{Sharafati et~al.(2021)Sharafati, {Haji Seyed Asadollah}, and Al-Ansari}]{SHARAFATI20213521}
\bibinfo{author}{A.~Sharafati}, \bibinfo{author}{S.~B. {Haji Seyed Asadollah}}, \bibinfo{author}{N.~Al-Ansari},
\newblock \bibinfo{title}{Application of bagging ensemble model for predicting compressive strength of hollow concrete masonry prism},
\newblock \bibinfo{journal}{Ain Shams Engineering Journal} \bibinfo{volume}{12} (\bibinfo{year}{2021}) \bibinfo{pages}{3521--3530}. \URLprefix \url{https://www.sciencedirect.com/science/article/pii/S2090447921002070}. \DOIprefix\doi{https://doi.org/10.1016/j.asej.2021.03.028}.
\bibitem[{Qian et~al.(2024)Qian, Ying, Lam, and Yin}]{qian2024baggingimprovesgeneralizationexponentially}
\bibinfo{author}{H.~Qian}, \bibinfo{author}{D.~Ying}, \bibinfo{author}{H.~Lam}, \bibinfo{author}{W.~Yin}, \bibinfo{title}{Bagging improves generalization exponentially}, \bibinfo{year}{2024}. \URLprefix \url{https://arxiv.org/abs/2405.14741}. \href{http://arxiv.org/abs/2405.14741}{{\tt arXiv:2405.14741}}.
\bibitem[{Van~der Maaten and Hinton(2008)}]{Maaten:Hinton:2008}
\bibinfo{author}{L.~Van~der Maaten}, \bibinfo{author}{G.~Hinton},
\newblock \bibinfo{title}{Visualizing data using t-sne.},
\newblock \bibinfo{journal}{Journal of machine learning research} \bibinfo{volume}{9} (\bibinfo{year}{2008}).
\bibitem[{Beltagy et~al.(2020)Beltagy, Peters, and Cohan}]{beltagy2020longformerlongdocumenttransformer}
\bibinfo{author}{I.~Beltagy}, \bibinfo{author}{M.~E. Peters}, \bibinfo{author}{A.~Cohan}, \bibinfo{title}{Longformer: The long-document transformer}, \bibinfo{year}{2020}. \URLprefix \url{https://arxiv.org/abs/2004.05150}. \href{http://arxiv.org/abs/2004.05150}{{\tt arXiv:2004.05150}}.
\bibitem[{Karpukhin et~al.(2020)Karpukhin, Oguz, Min, Lewis, Wu, Edunov, Chen, and Yih}]{Karpukhin:et:al:2020}
\bibinfo{author}{V.~Karpukhin}, \bibinfo{author}{B.~Oguz}, \bibinfo{author}{S.~Min}, \bibinfo{author}{P.~Lewis}, \bibinfo{author}{L.~Wu}, \bibinfo{author}{S.~Edunov}, \bibinfo{author}{D.~Chen}, \bibinfo{author}{W.-t. Yih},
\newblock \bibinfo{title}{Dense passage retrieval for open-domain question answering},
\newblock in: \bibinfo{editor}{B.~Webber}, \bibinfo{editor}{T.~Cohn}, \bibinfo{editor}{Y.~He}, \bibinfo{editor}{Y.~Liu} (Eds.), \bibinfo{booktitle}{Proceedings of the 2020 Conference on Empirical Methods in Natural Language Processing (EMNLP)}, \bibinfo{publisher}{Association for Computational Linguistics}, \bibinfo{address}{Online}, \bibinfo{year}{2020}, pp. \bibinfo{pages}{6769--6781}. \URLprefix \url{https://aclanthology.org/2020.emnlp-main.550}. \DOIprefix\doi{10.18653/v1/2020.emnlp-main.550}.
\bibitem[{Zhu et~al.(2024)Zhu, Yuan, Wang, Liu, Liu, Deng, Chen, Dou, and Wen}]{Zhu:Yuan:et:al:2024}
\bibinfo{author}{Y.~Zhu}, \bibinfo{author}{H.~Yuan}, \bibinfo{author}{S.~Wang}, \bibinfo{author}{J.~Liu}, \bibinfo{author}{W.~Liu}, \bibinfo{author}{C.~Deng}, \bibinfo{author}{H.~Chen}, \bibinfo{author}{Z.~Dou}, \bibinfo{author}{J.-R. Wen}, \bibinfo{title}{Large language models for information retrieval: A survey}, \bibinfo{year}{2024}. \URLprefix \url{https://arxiv.org/abs/2308.07107}. \href{http://arxiv.org/abs/2308.07107}{{\tt arXiv:2308.07107}}.
\bibitem[{Tsuda and Imaizumi(2023)}]{tsuda2023benignoverfittingnonsparsehighdimensional}
\bibinfo{author}{T.~Tsuda}, \bibinfo{author}{M.~Imaizumi}, \bibinfo{title}{Benign overfitting of non-sparse high-dimensional linear regression with correlated noise}, \bibinfo{year}{2023}. \URLprefix \url{https://arxiv.org/abs/2304.04037}. \href{http://arxiv.org/abs/2304.04037}{{\tt arXiv:2304.04037}}.
\bibitem[{Ng(2004)}]{l1l2}
\bibinfo{author}{A.~Y. Ng},
\newblock \bibinfo{title}{Feature selection, l1 vs. l2 regularization, and rotational invariance},
\newblock in: \bibinfo{booktitle}{Proceedings of the Twenty-First International Conference on Machine Learning}, ICML '04, \bibinfo{publisher}{Association for Computing Machinery}, \bibinfo{address}{New York, NY, USA}, \bibinfo{year}{2004}, p.~\bibinfo{pages}{78}. \URLprefix \url{https://doi.org/10.1145/1015330.1015435}. \DOIprefix\doi{10.1145/1015330.1015435}.
\bibitem[{Ghosh et~al.(2019)Ghosh, Dasgupta, and Swetapadma}]{svrsurvey}
\bibinfo{author}{S.~Ghosh}, \bibinfo{author}{A.~Dasgupta}, \bibinfo{author}{A.~Swetapadma},
\newblock \bibinfo{title}{A study on support vector machine based linear and non-linear pattern classification},
\newblock in: \bibinfo{booktitle}{2019 International Conference on Intelligent Sustainable Systems (ICISS)}, \bibinfo{year}{2019}, pp. \bibinfo{pages}{24--28}. \DOIprefix\doi{10.1109/ISS1.2019.8908018}.

\end{thebibliography}
